\documentclass[manuscript,screen,nonacm]{acmart} 
\acmConference[]{} 

\newcommand{\workshopname}{GenAICHI: CHI 2024 Workshop on Generative AI and HCI}
\newcommand{\licensedetails}{Licensed under a Creative Commons Attribution 4.0 International License (CC BY 4.0). Copyright remains with the author(s).}
\newcommand\extrafootertext[1]{
    \bgroup
    \renewcommand\thefootnote{\fnsymbol{footnote}}%
    \renewcommand\thempfootnote{\fnsymbol{mpfootnote}}%
    \footnotetext[0]{#1}%
    \egroup
}

\AtBeginDocument{ 
    \fancypagestyle{firstpagestyle}{
        \fancyhf{}
        \fancyfoot[L]{\sffamily\footnotesize \workshopname}%
        \fancyfoot[C]{\sffamily\footnotesize \thepage}
    }
    \fancyhf{}
    \fancyhead[L]{\sffamily\footnotesize\shorttitle}
    \fancyhead[R]{\sffamily\footnotesize\shortauthors}
    \fancyfoot[L]{\sffamily\footnotesize\workshopname}%
    \fancyfoot[C]{\sffamily\footnotesize\thepage}
    \extrafootertext{\licensedetails}
}

\usepackage{graphicx}                  
\usepackage{siunitx}                   
\usepackage{dblfloatfix}               
\usepackage{multimedia}                

\graphicspath{{figures/}{pictures/}{images/}{./}} 

\AtBeginDocument{%
  \providecommand\BibTeX{{%
    \normalfont B\kern-0.5em{\scshape i\kern-0.25em b}\kern-0.8em\TeX}}}

\begin{document}

\title[Leveraging AI to Generate Audio for User-generated Content in Video Games]{Leveraging AI to Generate Audio for User-generated Content in Video Games}

\author{Thomas Marrinan}
\email{tmarrinan@stthomas.edu}
\affiliation{%
  \institution{University of St. Thomas}
  \city{St. Paul}
  \state{Minnesota}
  \country{USA}
  \postcode{55105}
}
\affiliation{%
  \institution{Argonne National Laboratory}
  \city{Lemont}
  \state{Illinois}
  \country{USA}
  \postcode{60439}
}

\author{Pakeeza Akram}
\email{akra6203@stthomas.edu}
\affiliation{%
  \institution{University of St. Thomas}
  \city{St. Paul}
  \state{Minnesota}
  \country{USA}
  \postcode{55105}
}

\author{Oli Gurmessa}
\email{gurm1658@stthomas.edu}
\affiliation{%
  \institution{University of St. Thomas}
  \city{St. Paul}
  \state{Minnesota}
  \country{USA}
  \postcode{55105}
}

\author{Anthony Shishkin}
\email{shis4133@stthomas.edu}
\affiliation{%
  \institution{University of St. Thomas}
  \city{St. Paul}
  \state{Minnesota}
  \country{USA}
  \postcode{55105}
}

\renewcommand{\shortauthors}{Marrinan et al.}

\begin{abstract}
In video game design, audio (both environmental background music and object sound effects) play a critical role. Sounds are typically pre-created assets designed for specific locations or objects in a game. However, user-generated content is becoming increasingly popular in modern games (e.g. building custom environments or crafting unique objects). Since the possibilities are virtually limitless, it is impossible for game creators to pre-create audio for user-generated content. We explore the use of generative artificial intelligence to create music and sound effects on-the-fly based on user-generated content. We investigate two avenues for audio generation: 1) text-to-audio: using a text description of user-generated content as input to the audio generator, and 2) image-to-audio: using a rendering of the created environment or object as input to an image-to-text generator, then piping the resulting text description into the audio generator. In this paper we discuss ethical implications of using generative artificial intelligence for user-generated content and highlight two prototype games where audio is generated for user-created environments and objects.
\end{abstract}





\maketitle

\section{Introduction}
User-generated content (UGC) within digital platforms, notably in gaming and virtual environments, often grapples with significant challenges concerning audio integration \cite{Choi2016}. While users can readily create and customize visual elements like objects and characters, incorporating high-quality and dynamic audio remains a considerable obstacle \cite{Kafai2020}. Traditional methods of audio creation demand specialized skills and time-intensive processes and offer limited flexibility, leading to an imbalance between the visual and auditory aspects of user-generated content.

Addressing the audio challenges in UGC necessitates multifaceted solutions. One possible avenue is to develop simplified audio creation tools, offering intuitive interfaces that require minimal expertise \cite{Mitra2018}. Such tools would empower users to generate custom audio content without extensive training, thus democratizing audio creation. A second option could be to integrate dynamic audio systems that can adjust and adapt audio output in real-time based on factors such as user preferences, environmental conditions, etc. into content creation platforms \cite{Tan2019}. Lastly, the integration of generative artificial intelligence (AI) presents a promising avenue for automating and enhancing the audio generation process \cite{Elbamby2020}. By leveraging advanced AI algorithms, users can access scalable and customizable audio solutions that meet the diverse needs of their projects.

Generative AI technologies offer unique advantages in addressing the audio challenges for UGC. These algorithms provide unparalleled flexibility and adaptability, capable of producing diverse and dynamic audio content tailored to specific project requirements \cite{Hao2018}. Additionally, generative AI solutions are scalable, enabling the rapid generation of vast quantities of high-quality audio content \cite{Ding2021}. Moreover, the automation provided by generative AI reduces the burden on users, eliminating the need for extensive manual intervention and specialized skills \cite{Chen2019}. By continually learning from user feedback, generative AI models can ensure ongoing improvements in the quality and diversity of generated audio content, enhancing the overall user experience in user-generated environments.

\section{Ethical Considerations}
There are a number of ethical considerations when using generative AI. Training datasets that are created without diverse inputs or proper licensing can lead to output that may be biased or violate copyrights. Additionally, art (including music) is considered one of the defining characteristics of humanity \cite{MorrissKay_2010}. Therefore, a distinction must be made between humans using AI to assist in the creation of art and the use of generative AI to replace humans when creating art.

In this work, we selected MusicGen and AudioGen from Meta's AudioCraft \cite{AudioCraft} for generating music and sound. According to their documentation, ``MusicGen, which was \textbf{trained with Meta-owned and specifically licensed music}, generates music from text-based user inputs, while AudioGen, \textbf{trained on public sound effects}, generates audio from text-based user inputs.'' While we cannot be sure that the training datasets are unbiased, this does alleviate issues surrounding stolen or misused audio.

For our proposed workflow of leveraging generative AI to create audio for user-generated content, the AI accomplishes a task that is not feasible for humans alone. Expert audio creators are not available on demand and cannot create new sound effects or music tracks in mere seconds. However, we are not proposing to replace audio creators with generative AI either. Rather, we envision audio creators using generative AI as a tool -- enabling the software to base audio off of their expertly created music and sound effects. This could enhance the content they create and introduce new methods for listeners to interact with their work.

\section{Prototype Games}
Consider platforms like Roblox \cite{Roblox} and Dreams \cite{Dreams}, where players can design and share their game worlds, complete with custom assets, mechanics, and narratives. While these platforms offer extensive tools for level design and character customization, the ability to integrate dynamic music and sound into user-generated content remains relatively untapped.

In order to integrate custom audio for user-generated content, we leveraged Meta's AudioCraft \cite{AudioCraft}, a library that uses generative AI to create audio from text descriptions. We present two prototype games that generate audio based on user-generated content. The first game enables users to create a custom level in a simple two-dimensional (2D) platform game. We then use generative AI to create background music that suites the mood of the level the user created. The second 2D game enables users to create a custom vehicle that must cross a rough terrain. We then use generative AI to create sound effects for the custom vehicle.

\subsection{Game 1: User-Generated Environments}
\begin{figure*}[tb]
 \centering 
 \includegraphics[width=0.33\textwidth]{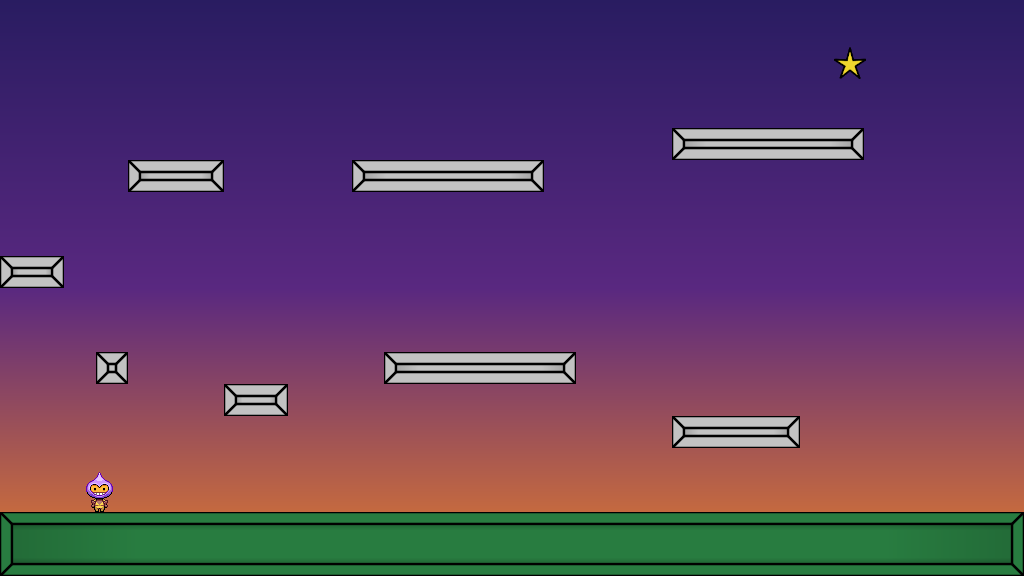}
 \includegraphics[width=0.33\textwidth]{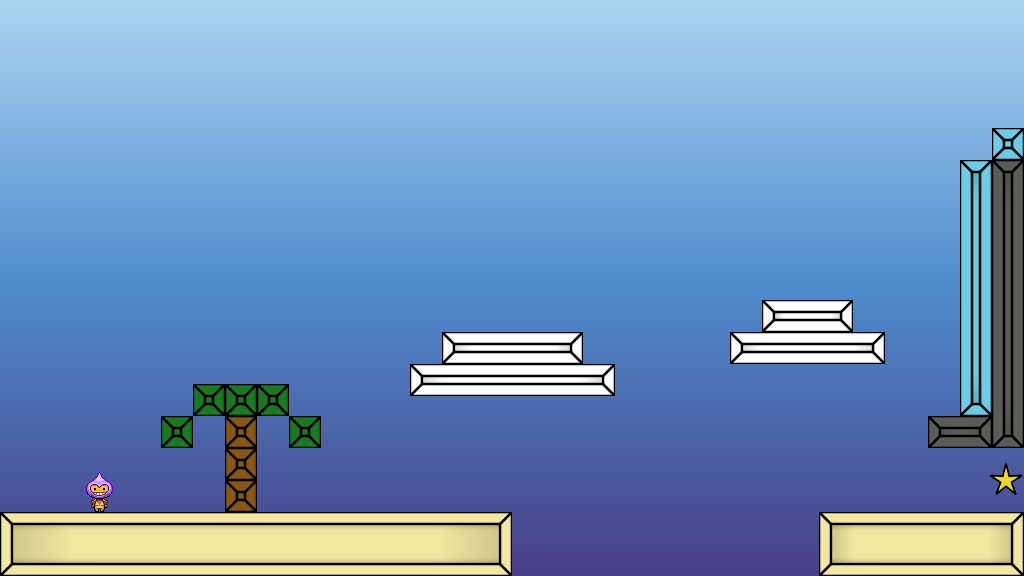}
 \includegraphics[width=0.33\textwidth]{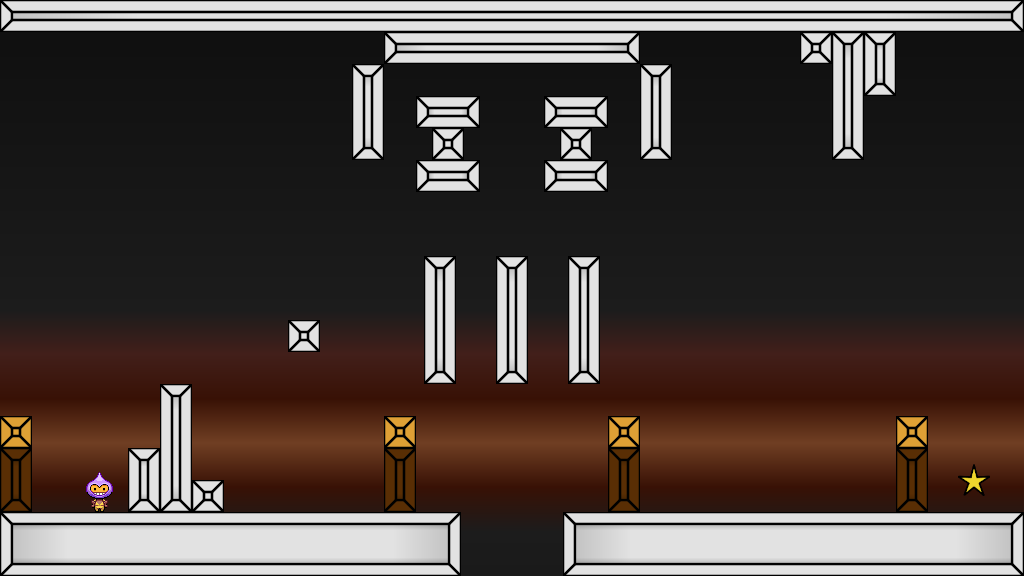}
 \caption{Three pre-created levels. In each level, the user controls a character to jump between platforms and reach the star. Each level has its own background music appropriate to its aesthetics (grasslands at twilight, tropical beach, and ominous cave).}
 \label{fig:game1-levels}
\end{figure*}

\begin{figure*}[b]
 \centering 
 \includegraphics[width=0.4473\textwidth]{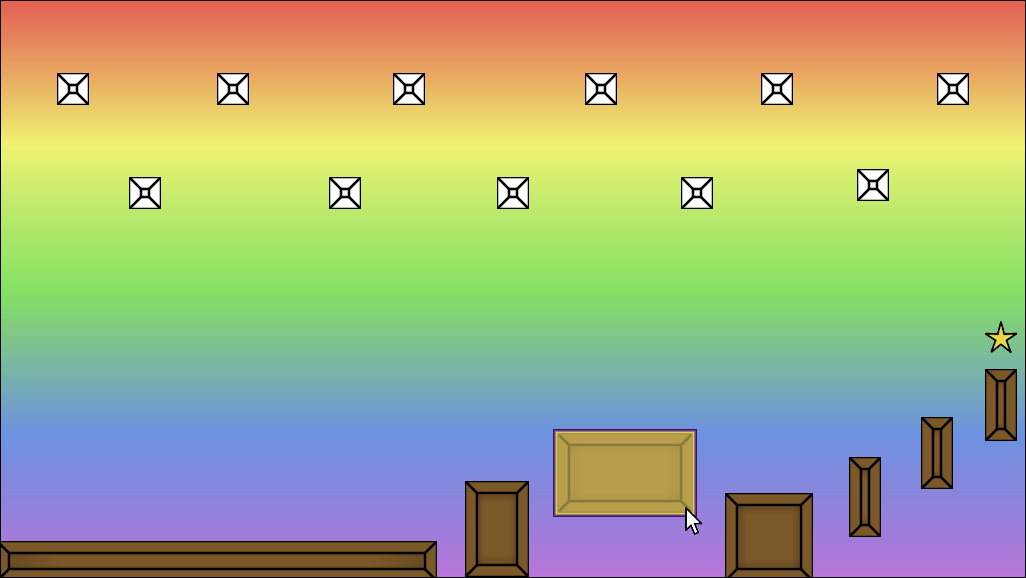}
 \includegraphics[width=0.3644\textwidth]{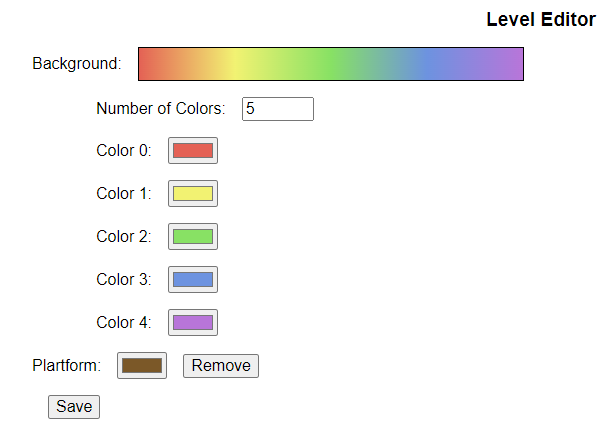}
 \caption{Level editor. Users can select colors for making a background gradient, add a new platform, move an existing platform, change the color of an existing platform, and select the position of the goal.}
 \label{fig:game1-editor}
\end{figure*}

We created a prototype 2D game where the user controls a character jumping between platforms to reach a goal. Our prototype includes three pre-created levels (shown in \autoref{fig:game1-levels}) and a level creator mode. The level creator enables users to design their own level by setting the background gradient, adding and editing platforms, and choosing the location of the goal. Once the user is satisfied with their level, they can save and play it. An example of a user creating their own custom level is shown in \autoref{fig:game1-editor}.

The three pre-created levels each have environmental audio tracks while being played. To create an audio track for a user-generated level, we leveraged MusicGen \cite{AudioCraft} to dynamically generate background music based on a text description. We investigated two methods for programmatically creating a text description for a user-generated level: 1) use colors in the background gradient to create a mood description, and 2) use an image of the user-generated level as input to BLIP \cite{Li_2022} (an AI image captioning technique) to generates a text description for the image. 

To create a text description based on the user-generated background gradient, we compare each color in the gradient to a set of predefined colors in order to determine which it is most similar to. We also chose a descriptive word that represents a mood commonly associated with each of the predefined colors. If four or more unique colors are present (excluding black, white, and brown), we consider the background to be a rainbow and select the mood ``playful''. Otherwise, we take one or two colors that occur most frequently in the gradient and use their moods. The colors and their associated moods we use are: red -- ``intense'', orange -- ``energetic'', yellow -- ``cheery'', green -- ``fresh'', cyan -- ``lively'', blue -- ``peaceful'', purple -- ``mysterious'', pink -- ``cute'', brown -- ``practical'', black -- ``dark'', and white -- ``simple''.

To programmatically create a prompt for MusicGen to generate background music, we plugged the mood descriptor into the phrase: ``90s game vibe with \_\_\_\_\_ chiptunes and 8-bit melodies.'' Alternatively, we also tested creating the prompt for MusicGen by using BLIP to generate a text description from an image of the user-created level. In either case, the prompts were used to generate 8-second audio clips. On our test machine (AMD Threadripper 2950X CPU and NVIDIA RTX 2080 GPU), these clips took less than 4 seconds to generate, thus demonstrating responsiveness in real-time user-generated content scenarios. 



Furthermore, MusicGen supports a multi-modal approach to input, accepting prompts that combine text with existing audio. This functionality generates new music, using the melody and sample rate of the input audio, and fitting other musical properties to the text description. This can be useful in future applications, for example, if a game's audio creator were to develop baseline theme music that the AI could then tweak to fit various moods.

\subsection{Game 2: User-Generated Objects}
\begin{figure*}[tb]
 \centering 
 \includegraphics[width=0.41\textwidth]{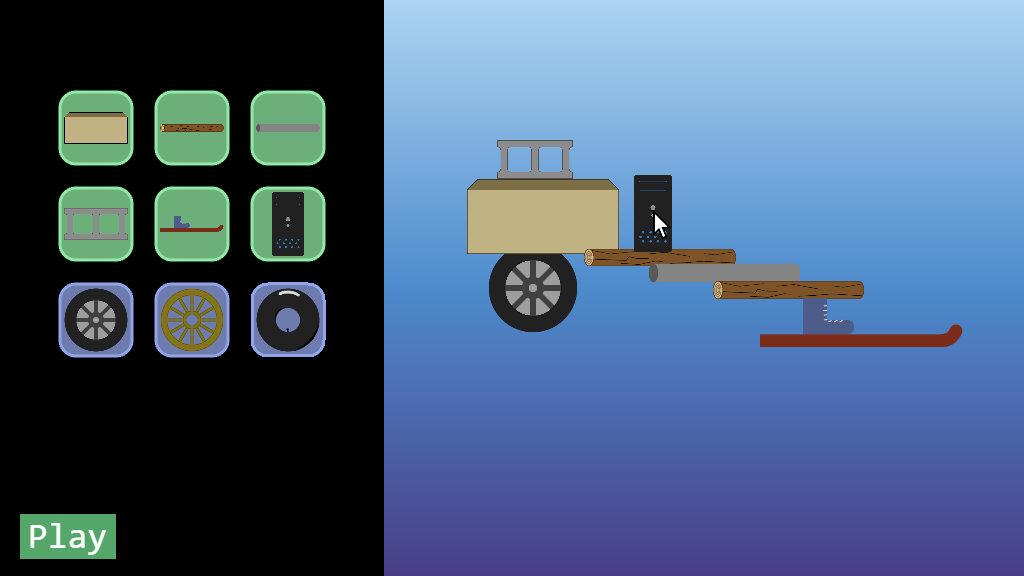}
 \includegraphics[width=0.41\textwidth]{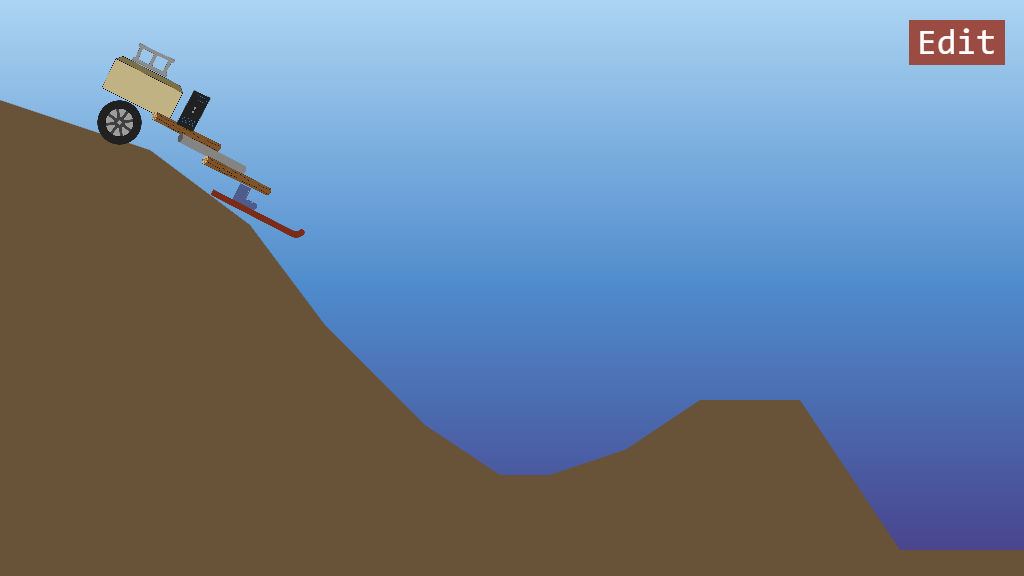}
 \caption{Editing and testing a user-generated vehicle. Left panel shows the vehicle editor, where users can drag components around to create a custom vehicle. Right panel shows the completed vehicle attempting to cross the rough terrain.}
 \label{fig:game2}
\end{figure*}

We created a prototype 2D game where users drive a vehicle across rough terrain to reach the finish line. First, users must build their vehicle, using a set of predefined components. Then they test to see if their vehicle can cross the rough terrain or if it gets stuck. Snapshots from these two phases of the gameplay are shown in \autoref{fig:game2}. Since each vehicle is unique, we leveraged AudioGen \cite{AudioGen} to dynamically generate an audio sound effect based on a text description. Similar to the first game, we investigated two methods for programmatically creating a text description for a user-generated vehicle: 1) use the components of the vehicle to create a description, and 2) use an image of the user-generated vehicle as input to BLIP \cite{Li_2022} to generate a text description for the image.

To build a vehicle, users have three choices for wheels -- wooden wagon wheel, rubber car tire, and inflatable inner tube. Users build the vehicle body from an assortment of other objects -- cardboard boxes, skis, cinder blocks, steel pipe, etc. Each wheel has its own properties for mass, friction, and bounciness that impact how well the vehicle will be able to cross the terrain. Similarly, each object that makes up the vehicle body has its own mass and friction that also impacts the vehicle's ability to cross the terrain. Users are able to create their vehicle using any number of each object.

During the testing phase, users watch as their created vehicle attempts to drive over the terrain and reach the finish line. To simulate driving a powered vehicle, the angular velocity of the wheels accelerate until reaching a maximum speed, thus propelling the vehicle forward when the wheels are in contact with the ground. If this force is insufficient to overcome the friction and gravity associated with the terrain, users can go back to the editor and modify their vehicle before testing again.

Once users create their vehicle, but before testing it, we generate a sound effect for that vehicle. To programmatically create a prompt for AudioGen to generate a sound effect, we used the overall vehicle mass and descriptions of its wheels; for example: ``light vehicle with wooden wheels.'' Alternatively, we also tested creating the prompt for AudioGen by using BLIP to generate a text description from an image of the user-created vehicle with the constraint that the prompt must begin with the word ``vehicle''. In either case, the prompts were used to generate 5-second audio clips. Similar to MusicGen, these clips took less than 4 seconds to generate on our test machine (AMD Threadripper 2950X CPU and NVIDIA RTX 2080 GPU), thus demonstrating responsiveness in real-time user-generated content scenarios.

\section{Discussion}
Overall, the audio created by MusicGen and AudioGen was of high-quality and fit the aesthetics of the 2D games. For Game 1, while the programmatic text description and image-to-text description resulted in different background music, we felt both were extremely good. Similarly for Game 2, the the programmatic text descriptions and image-to-text descriptions resulted in differing sound effects. However, both successfully generated audio that fit the general idea of a moving vehicle and enhanced the overall experience of the game. In all cases, the quality of the text prompt greatly impacted how good the resulting audio fits its use. It took significant tweaking to finalize the music prompt to insert the mood descriptors into. While we generally felt that the sound effects did not match the game quite as well as the environmental music, we believe that more accurate audio could be generated with further testing and modification to the text descriptions of the vehicles.

Worth noting, is how impressed we were with the results of the image-to-text being used as prompts for generative audio. We were concerned that having multiple stages of generative AI could lead to more potential error. However, for both music and sound effects, the results were compelling.

Finally, we are excited to extend this work to better incorporate audio content creators during the generative stage. We would like to use the multi-modal text plus audio mode in MusicGen to tweak expertly created audio to fit moods or environments of user-generated content. We would also like to incorporate pre-created game audio into the AI training dataset to see if it is possible for it to generate music and sound effects in the style of specific games. These steps would keep audio creators as critical pieces of the video game design process and enable them to deliver their content in new and exciting ways.

\section{Conclusion}
In this paper, we aimed to investigate the technical capacities of generative AI for creating audio for user-generated content in video games. We motivated our work, noting the difficultly in rapidly creating quality audio, and discussed some ethical implications of using generative AI. We then showcased two prototype games that leverage generative AI, one to create environmental background music and another to create object sound effects. For both games, we examined the capabilities of programmatically constructing text descriptions as well as using AI for creating text descriptions from an image. These text descriptions became the input prompts for audio generation.

In our qualitative assessment, we felt that AI generated audio that satisfactorily fit with the game aesthetics. And most importantly, the rapid creation of quality audio enabled by generative AI opened up avenues previously inaccessible for user-generated content. A demonstration of both games and sample generated audio is available at \url{https://youtu.be/6eSGYm7v3VA}.

In the future, we plan to further investigate various aspects of AI assisted audio generation. To make generated audio feel more like it belongs to the specific game, we are interested in exploring two avenues. First, we would like to tweak generated audio to gain stylistic similarities to pre-created game audio. Second, we would like to expand the training datasets to include expertly created audio tagged for various styles of video games. Additionally, we are interested in looking into human-in-the-loop audio generation. By providing users the text prompt used to generate the audio, they could be given an opportunity to iteratively edit the prompt and listen to the resulting audio until satisfied. It is our belief that the inclusion of rapid, high-quality audio into user-generated content systems will enhance player creativity, community engagement, and overall gaming experience.

\begin{acks}
We would like to thank the University of St. Thomas Undergraduate Research Opportunities Program and the Department of Computer and Information Sciences for their support in presenting this work.
This research was also supported in part by the Argonne Leadership Computing Facility, which is a U.S. Department of Energy Office of Science User Facility operated under contract DE-AC02-06CH11357.
\end{acks}

\bibliographystyle{ACM-Reference-Format}
\bibliography{references}

\end{document}